# Superdislocations and point defects in pyrochlore Yb$_2$Ti$_2$O$_7$ single crystals and implication on magnetic ground states

Zahra Shafieizadeh[1,+], Yan Xin[2,+,*], Seyed M. Koohpayeh[3], Qing Huang[4], & Haidong Zhou[4]

[1] Department of Physics, Florida State University, Tallahassee, FL 32311, USA

[2] National High Magnetic Field Laboratory, Florida State University, Tallahassee, FL 32310, USA

[3] Institute for Quantum Matter, Department of Physics and Astronomy, Johns Hopkins University, Baltimore, MD 21218, USA

[4] Department of Physics and Astronomy, University of Tennessee, Knoxville, TN 37996, USA

[+] These two authors contributed equally to this work.

[*]Correspondence and requests for materials should be addressed to Y.X. (email: xin@magnet.fsu.edu)

Contact information:

Dr. Yan Xin

National High Magnetic Field Laboratory

Florida State University

1800 E. Paul Dirac Drive, Tallahassee, FL 32310

USA

Phone: (850) 644-1529

e-mail: xin@magnet.fsu.edu

**Abstract**

This study reports atomic-scale characterization of structural defects in $Yb_2Ti_2O_7$, a pyrochlore oxide whose subtle magnetic interactions is prone to small perturbations. Due to discrepancies in the reported magnetic ground states, it has become a pressing issue to determine the nature of defects in this system. In the present study, we use atomic resolution scanning transmission electron microscopy techniques to identify the type of defects in the ytterbium titanate single crystals grown by the conventional optical floating zone (FZ) method. In addition to the known point defects of substitution Yb on Ti B-sites, extended defects such as dissociated superdislocations and anti-phase boundaries were discovered for the first time in this material. Such defects were prevalently observed in the FZ grown single crystals (of a darker color), in contrast to the stoichiometric white polycrystalline powders or high quality colorless single crystals grown by the traveling solvent floating zone (TSFZ) technique. The lattice strains from these extended defects result in distortions of Yb-tetrahedron. A change of Ti valance was not detected at the defects. Our findings provide new insights into understanding the nature of defects that are of great importance for the physical property studies of geometrically frustrated compounds. Furthermore, this work sheds light on the complicated core structure of superdislocations that have large Burgers vectors in oxides with complex unit cells.

## Introduction

Pyrochlore oxides $A_2B_2O_7$, particularly $RE_2Ti_2O_7$ ($RE$ = Yb, Ho, Tb and Dy) has been of great interests due to their geometrically frustrated magnetic interactions, displaying different magnetic ground states arising from the interplay between geometry frustration and spin orbital interaction, such as spin-ice, spin-liquid, long-range, or short-range ferromagnetic ordering[1-5]. For $Yb_2Ti_2O_7$, high sensitivity to small perturbations of all kinds is known to significantly affect the magnetic ground states. This compound, as a quantum spin liquid candidate, is poised to be at the boundaries of ferromagnetic and antiferromagnetic states, and the reported physical properties vary from sample to sample[4-10]. Although the single crystal form of the $Yb_2Ti_2O_7$ compound can be readily grown by the conventional FZ method, the quality of the crystals varies considerably, e.g. showing broader specific heat features at 150 to 200 mK and different magnetic ground states. This is in contrast with a sharp peak observed at 250 mK in the stoichiometric polycrystalline samples synthesized by a solid-state reaction or the high-quality crystals grown by the traveling solvent floating zone technique (TSFZ)[5-9]. The work on the effect of doping on its physical properties has also supported the idea that the FZ single crystals are non-stoichiometric[9].

In a recent neutron scattering study, a "stuffed" structure was proposed, in which about 2.3% excess Yb atoms substitute B-site Ti atoms, while the possibility of other point defects such as Yb and Ti anti-site disorders or Ti vacancies was also suggested[10]. In another study, substitution Yb on Ti B-site was observed by atomic resolution energy dispersive X-ray spectroscopy[11]. Furthermore, extended X-ray absorption fine structure (EXAFS) at the $Yb^{3+}$ $L_3$-edge indicated a higher level of disorder or the possible presence of vacancies[4]. Using diffuse neutron scattering and magnetization measurements, the important role of oxygen vacancies in the low temperature monopole physics of as grown $Y_2Ti_2O_7$ crystals has also been reported[12].

Following these studies, the general consensus is that non-stoichiometry and point defects (such as stuffing or oxygen vacancies) are mainly responsible for varied magnetic properties.

In this report, we used atomic resolution high-angle-annular-dark-field scanning transmission electron microscopy (HAADF-STEM) imaging and electron energy loss spectroscopy (EELS) to characterize the structural defects in $Yb_2Ti_2O_7$ single crystals. Three sample types were used for this study, the stoichiometrically synthesized powder and the FZ crystal (both samples from University of Tennessee), together with single crystals grown by the FZ and TSFZ techniques (from Johns Hopkins University). Our results revealed substantial difference in defect density among the samples. In contrast to both high quality samples of the stoichiometric powder and the TSFZ single crystal[9], high defect densities of a different nature were detected in the low quality FZ crystals. For the first time, extended defects such as dissociated superdislocations (with Burgers vector size larger than 0.6 nm) and anti-phase boundaries (APBs) are observed, in addition to the known point defects of substitution Yb atoms on Ti B-sites. The existence of these extended defects induces considerable strain and such strain causes the distortion of Yb-tetrahedron. This potentially disrupts long-range structural perfection of the material, which would ultimately disturb the delicate magnetic interactions.

Dislocations are important line defects for materials as they are not only essential in materials' plasticity, but also affect physical properties such as acting as diffusion route[13], influence on device performance[14], or enhancing superconductivity[15]. In this study, the core structure of superdislocations with large Burgers vector seems to be different, in comparison to the core structures of dislocations in oxide materials such as in $SrTiO_3$[16], $YBa_2Cu_3O_7$[17,18], α-$Al_2O_3$[19], $NiO$[20] or at oxide interface[21]. Size of Burgers vector reported in the past is usually in the range from 0.159 nm for α-$Al_2O_3$ to 0.421 nm for MgO, while it is larger than 0.6 nm in the pyrochlore material.

## Results

The samples used in this study were well characterized by powder X-ray diffraction and magnetization measurements, and some details were reported in an earlier work[9]. It was determined that all samples have a cubic pyrochlore structure with space group $Fd\bar{3}m$. The lattice parameter measured for the FZ single crystal ($a = 10.03853(5)$ Å) was about 0.083% larger than the stoichiometric powder ($a$ = 10.03023(5) Å) or the TSFZ crystal ($a$ = 10.03066(6) Å)[9]. A larger lattice constant for the FZ grown crystal is consistent with larger $Yb^{3+}$ replacing smaller $Ti^{4+}$ cations[9]. Physical property measurements of similar quality samples have also shown that the peak in the low-temperature heat capacity measurements taken from the FZ crystals shifts to lower temperatures and becomes weaker and broader[6], as similarly seen for the Ti-deficient stuffed powders[9]. For the closely stoichiometric powders and the TSFZ single crystals, however, a sharp peak was obtained for the specific heat data[6, 9]. The Curie-Weiss temperature is 0.39 K, lower than the polycrystalline sample of 0.82 K (University of Tennessee samples) or TSFZ crystals of 0.84 K (Johns Hopkins University sample)[9].

Atomic resolution HAADF-STEM imaging is a powerful tool to observe and obtain information on defect structures due to the relatively simple relationship of the atomic column image intensity to atomic number Z and sample thickness[22]. To study the ytterbium titanate crystals, the best choice of crystal orientation is along <110> where there are atomic columns only consisting of Yb or Ti atoms. The images were taken under the experimental conditions that the atomic column intensity difference between Yb and Ti columns is mainly from the atomic number difference with minimum diffraction contrast effect (Supplementary Information, Figure S1). Signals from oxygen atomic columns are too low to be detected at this imaging condition, therefore, for the images presented in this work, the lowest intensity columns are the Ti atoms (Z = 22), while the columns of mixed Ti and Yb atoms (Z = 46) show medium bright intensity, and the column containing only Yb atoms (Z = 70) has highest intensity.

We found that the high quality TSFZ crystal (Figure 1a-c) and the polycrystalline powder sample (Figure 1d-f) have a nearly perfect structure, compared to the FZ single crystals in which point defects were seen (Figure 2). The fractured crystal pieces used for STEM observation usually have thickness variation across only a few nanometers; therefore, we used Absolute Integrator image processing software[23] to map and survey the atomic column intensity, which provided an overview of the atomic column thickness variation across the entire image (Figure 1b and 1e). Yb column intensity mainly changes with column thickness, and it does not have appreciable intensity change even if there are anti-site disorders of Ti goes into Yb column, i.e. when some Ti substitute into Yb position (Supplementary Information, Figure S2d). Therefore, the intensity from a column of only Yb atoms can be used as a reference for determining adjacent Ti atomic column thickness.

Yb columns in both crystals along [110] with comparable intensity were selected, as indicated by rectangular boxes in the original images (Figures 1a and 1d). The Yb column intensity was measured by integrating intensity from an area of 13×13 pixels, and each individual Ti column intensity was obtained by integrating the pixel intensity from an area of 8×8 pixels. Atomic column intensities in red boxes are plotted in Figure 1c and 1f. This method of integrated scattering cross-section of atomic column is practically useful in measuring column intensity for quantitative comparison[24]. The intensity variation of Yb column is about 3.8%, which could arise from statistic error, signal noise, random vacancies in atomic columns, sample surface amorphous layer or electronic noise. Ti atomic columns in between these Yb columns have similar column thickness for the plotted atoms. The blue bar graphs in Figure 1c and 1f show intensity variations of Ti atomic columns with similar thickness. For a perfect lattice, Ti column intensity fluctuates about 5.2%. Substitution Yb point defects, i.e. "stuffing", were less frequently observed and fewer in numbers in polycrystalline sample and in TFSZ single crystals.

However, in the colored FZ single-crystal samples, Yb "stuffed" Ti columns were frequently observed, as shown in Figure 2. Based on binominal distribution of the calculated probability of finding Yb atoms in a Ti column, it was suggested that Ti columns with "stuffed" Yb atom should be quite common even only for a few percentage[11]. We semi-quantitatively compare Ti column intensities of the same thickness to ascertain the Yb "stuffed" Ti Columns, and this comparison method is quite widely used[25, 26]. From the intensity map processed by Absolute Integrator (Figure 2b), the sample thickness varies noticeably within a few columns. Nevertheless we could select atomic columns that have similar thickness. To illustrate the Yb "stuffed" Ti columns, we selected four atomic rows as indicated in Figure 2a. For row 1, even with lower sample thickness indicated by the lower Yb column intensity (Yb column 5 and 6 in Figure 2c), the adjacent two Ti columns indicated by arrows have an intensity 41% and 46% higher than the average, which was based on the three thicker Ti columns. For row 2, the Yb columns have comparable thickness, but one Ti column intensity is obviously much brighter, 127% higher than average. For row 3, the two brighter Ti columns have 75% and 56% higher intensities than the adjacent Ti. For row 4, one Ti column is 84% brighter than the adjacent Ti columns. Since these brighter Ti columns show much higher intensity than the statistical variation of typical 5% uncertainty, we can conclude that they are Ti columns with substitution Yb atoms in them.

To further support this conclusion, we calculated the Ti column intensity vs thickness curve with one to five Yb atoms at different depth along the Ti column using xHREM[TM] STEM image simulation software (Supplementary Information, Figure S2a-c). The calculation showed that Ti column intensity increases rather discernibly when substitution Yb atoms exist in a Ti column below sample surface. Even just one substitution Yb atom can make the intensity more than 15% higher than the pure Ti column. Furthermore, the column intensity increases similarly whether the Yb atoms are segregated or randomly distributed along the Ti Column (Supplementary Information, Figure S2c). Based on these calculations and experimental results, it is clear that Ti atoms are replaced by some Yb atoms when a Ti column shows higher intensity than other Ti columns of the same thickness in the FZ single crystals. These "stuffed" Ti columns seem to be randomly distributed. As to anti-site disorders (substitution of Ti into Yb sites), it seems to be technically very difficult to detect them, since no appreciable intensity changes could be seen at the Yb column (Supplementary Information, Figure S2d). From the synthesis and crystal growth point of view, anti-stuffing is not expected to have happened in the FZ crystal since the grown crystal is Ti deficient and shows a larger lattice parameter[9].

In addition to the known point defects shown above, line and planar defects were observed in the FZ single-crystals. We have found two types of superdislocations with different Burgers vectors. One of the more frequently observed superdislocations in the FZ single-crystal samples is shown in Figure 3. The in-plane component of its Burgers vector projected along the electron beam direction onto the image plane can be determined by drawing a finish-start/right-hand (FS/RH) Burgers circuit around an end-on dislocation in an atomic resolution image. So by drawing a Burgers circuit around the whole defect in Figure 3a, we determined that this superdislocation has a projected Burgers vector of ¼[1$\bar{1}$2]. The inverse FFT images using $\mathbf{g}$ = ±(2$\bar{2}$2) and $\mathbf{g}$ = ± (2$\bar{2}\bar{2}$) diffractions respectively showed only one set of extra planes that are separated by 9 (1$\bar{1}$1) atomic planes, where {111} is the glide plane. Around the end of the extra planes, lay the cores of two partial dislocations, and they are on different glide planes separated by six (1$\bar{1}\bar{1}$) atomic planes. This superdislcoation dissociated and the partial dislocations climbed onto different glide plane. The two partial dislocations sandwich a {111} APB that is about eight atomic planes wide and 9 nm long. The details of the APB characteristics are shown later in Figure 4. The projected Burgers vectors for the partial dislocations are deduced to be $\mathbf{b}_1 = 1/8[1\bar{1}2]$ and $\mathbf{b}_2 = 1/8[1\bar{1}2]$. Since 1/8<112> is not a valid translational vector for this f.c.c structure, there must be a screw component perpendicular to the image plane. The core structure of the partial dislocations is quite complex and a typical core structure is clearly shown in Figure 3d. Its core is much more extended with a width of five $(1\bar{1}1)$ planes and a length of $(1\bar{1}\bar{1})$ eight planes. At the core, each atomic column appears to split into two columns along $[1\bar{1}2]$, which is formed by shifting the columns by $1/8[1\bar{1}2]$. This shift is obvious from inverse FFT image in Figure 4e, the atomic planes on the left side seem to have a kink relative to the right side of the core. Since it is energetically unfavorable to have two columns so close to each other, these closely positioned columns probably contain atoms shifted from each other along the out-of-plane direction. Therefore we speculate that there is a screw component of the Burgers vector of 1/8<110>. We conjecture that this superdislocation has a Burgers vector of $\frac{1}{4}[1\bar{1}2]$, and it dissociates into two partial dislocations with opposite signs of screw component, and an APB: $\mathbf{b} = \mathbf{b}_1 + \mathbf{b}_2$, where $\mathbf{b} = 1/4[1\bar{1}2]$, $\mathbf{b}_1 = 1/8[1\bar{1}2]_{edge} + 1/8[110]_{screw}$, and $\mathbf{b}_2 = 1/8[1\bar{1}2]_{edge} + 1/8[\bar{1}\bar{1}0]_{screw}$. This configuration seems to be reasonable, since it is consistent with the energy criterion for dislocation dissociation $\mathbf{b}^2 > \mathbf{b}_1^2 + \mathbf{b}_2^2$, and the Burgers vector is similar to the previously reported Burgers vector in a dislocation loop caused by electron irradiation[27].

To analyze strain field of the defects, we used geometric phase analysis (GPA), which is an image-processing routine to measure strain tensor, which utilizes small displacements of the lattice fringes at the defects relative to a reference lattice in perfect area of atomic resolution images[28]. GPA has been very useful in determining strain field of individual dislocations or defects using atomic resolution images[29]. For HAADF-STEM images, although the image contrast is less sensitive to local sample variation and image defocus, sample drifting and scan distortion would induce errors. The uncertainty of the measured strain is less than 1%[30, 31]. From our analysis, we estimate that there is up to about 2% uncertainty in the strain measurement. We used GPA DigitalMicrograph plug-in software developed by the HREM Research, Inc[32] to extract strain field map for the dissociated superdislocation from Figure 3a. In this case, the perfect area on the upper left corner of the image was used as a reference. $\varepsilon_{xx}$ (along $[1\bar{1}0]$) strain map calculated using GPA showed tensile strain about 6% to 10% above the extra-plane side of the cores for the upper partial dislocation and -2% strain (compressive) for the lower partial dislocation (Figure 3f). For $\varepsilon_{yy}$ along $[00\bar{1}]$, there is -3% compressive strain and 2% tensile strain respectively for the two partials. The strain from APB dominates the defect and it has about -10% compressive strain along $[1\bar{1}0]$, and about 10% tensile strain along $[00\bar{1}]$ (Figure 3g-h).

The dissociation of ¼<112> superdislocation generates APBs on {111} planes, and the atomic structure and characteristics of {111} APB are illustrated in Figure 4. The intensity line profiles of Ti-Yb/Ti and Yb/Ti rows in a perfect area show alternating high and low intensity on $(1\bar{1}\bar{1})$ planes along $[1\bar{1}2]$ (Figure 4a). However, little intensity difference was seen for the columns in rows in some areas, where the column intensities are similar (Figure 4b). This intensity feature is attributed to APB as shown more clearly in Figure 4c, where the Ti-Yb/Ti rows mismatched to the Yb/Ti rows on $(1\bar{1}1)$ plane across the APB. The phase image containing the APB was created by using $\mathbf{g} = (1\bar{1}1)$ (Figure 4d). It clearly confirmed the anti-phase nature of the boundary, i.e. the crystal to the right of the boundary has $-\pi$ phase and changes to $\pi$ across the boundary to the left side of the image. This APB phase change is more abrupt for most of the lower part, while it becomes a bit extended at the upper part (green colored area in Figure 4d), which indicates an extensive and complex APB shape. This APB lies on $(1\bar{1}\bar{1})$ plane and has a displacement of $1/8[1\bar{1}2]$. The lattice distortion, i.e. strain associated with APB can be calculated using $\mathbf{g} = (1\bar{1}\bar{1})$ and $\mathbf{g} = (1\bar{1}1)$ using GPA. The calculation showed that along $[1\bar{1}\bar{2}]$, APB experiences a compressive strain $\varepsilon_{xx}$ of about -4%, and a larger tensile strain $\varepsilon_{yy}$ of about 8% along $[1\bar{1}1]$. The shear strain is tensile in nature, and is on the order of about 6%.

A less frequently observed superdislocation has a larger Burgers vector of ½<110>. Figure 5 shows a dissociated dislocation partial of this type. From the Burgers vector circuit, we determined that this partial dislocation has an in-plane Burgers vector of ¼$[1\bar{1}0]$. A closer examination revealed that the atomic structure of the APB is three atomic layers thick lying on (001) planes instead of the frequently observed {111} type. This APB is bounded by one partial dislocation at the end. The inverse FFT images (Figure 5b and 5c) using $\pm(2\bar{2}2)$ and $\pm(\bar{2}22)$ diffractions show two {111} extra planes. In fact, these two extra planes meet at the partial dislocation core (Figure 5d). The partial dislocation core has a simple structure of a deformed hexagon shaped ring of about 1.1 nm in diameter. The anti-phase nature of this (001) APB is confirmed by the phase image in Fig. 5e. GPA strain field analysis of the (001) APB shows compressive strain of about -13% along $[1\bar{1}1]$ and a tensile strain of about 20% along $[1\bar{1}\bar{2}]$. The shear strain is compressive at about -3% (Figure 5f-i). It should be noted that (001) APB has higher strain energy than {111} APBs. The (001) APB has a displacement value of ¼$[1\bar{1}0]$. Since a pure edge dislocation in a f.c.c crystal structure usually has two extra planes end at the core, and the fact that this partial dislocation has two extra half-planes, we postulate that there is no screw component of the Burgers vector and the observed partial dislocation of b = ¼$[1\bar{1}0]$ might be dissociated from a pure edge dislocation of ½$[1\bar{1}0]$. GPA also showed that there is a compressive strain of -2% along $[1\bar{1}1]$ opposite the extra plane side of the partial dislocation core, and a tensile strain of 10% around the rest of the core area along $[1\bar{1}\bar{2}]$ (Figure 5f-g).

EELS in TEM has the advantage of studying electronic structures with near atomic spatial resolution. The electron energy loss near-edge structure (ELNES) of Ti $L_{2,3}$ EELS core-loss spectrum has been quite informative in determining Ti valence state and site geometry of Ti[33]. We collected EELS spectra from both perfect and APBs and made a comparison. Figure 6a shows superimposed two Ti $L_{2,3}$ spectra from perfect and APB areas. It consists of four major peaks, with $L_3$ (transitions from $2p_{3/2}$ to empty 3d orbital) having the first doublet at lower energy, and $L_2$ (transitions from $2p_{1/2}$ to empty 3d orbital) at higher energy. The doublet peaks are due to octahedral crystal field $t_{2g}$-$e_g$ splitting of 3d orbital. It has been found that there is a 1.7-2.0 eV shift to the lower energy side for the $L_{2,3}$ onset from Ti$^{4+}$ to Ti$^{3+}$. In addition, the relative height of the doublet peaks is sensitive to the Ti symmetry, i.e. the Ti-O octahedron distortion[33]. For comparison, the intensity of the spectra is normalized to the first peak of the Ti $L_3$ doublet. Using zero-loss peak as a reference, we found no $L_3$ onset energy shift within experimental uncertainty of 0.2 eV, but the peak height ratio of $L_2$ doublet at APBs is slightly larger. This ELNES change is consistent with previously reported[11]. However, we think that slight changes in peak area or height ratio alone are not sufficient to determine that Ti valence has changed from Ti$^{4+}$ to Ti$^{3+}$. This argument is supported by the fact that the ELNES of $L_2$ doublet peak ratio of rutile and anatase is different even though they both have Ti$^{4+}$. This peak height ratio difference seems to be more to do with Ti-O octahedron distortion[33]. Since APB has considerably large strain, the Ti-O octahedra are likely distorted in comparison to the perfect area, which results in the $L_{2,3}$ ELNES on APB has a slightly increased $L_2$ doublet peak ratio. There is no difference in O-K edge spectra ELNES between perfect and defective areas, as shown in Figure 6b. The first two peaks within 10 eV of the onset are the hybridization between oxygen 2p with metal 3d $t_{2g}$ and $e_g$ orbitals.

## Discussion and Summary

Using a direct imaging technique, the structural quality and defects were characterized for the ytterbium titanate pyrochlore crystals at the atomic scale. Besides the previously reported point defects (i.e. such as Yb substitution in Ti B-sites), we observed the extended planar and line defects, which are superdislocations and APBs. To the best of our knowledge, these types of defects were observed for the first time in this material. In contrast to the FZ single-crystal samples, we found far less defect densities in the stoichiometrically synthesized polycrystalline samples, while no defects of any types were seen in the high quality single crystal grown by the TSFZ. The fact that superdislcations and APBs were quite easily found in the FZ single crystals indicates their considerable density inside the crystals.

APBs are formed as a result of dissociation of superdislocations. The superdislocation with Burgers vector ¼<112> (0.631 nm) has lower energy than the ones with Burgers vector of ½<110> (0.729 nm), as dislocation energy is proportional to $\mathbf{b}^2$. Consequently, the ¼ <112> superdislocations are more frequently observed. All superdislocations dissociate into two partial dislocations, and an APB with several nanometers in width and length. The columns of dislocation cores have severe distortions, and the atoms at the core could either have dangling bonds or bonds that are highly distorted in both lengths and angles. The dislocations can only terminate at the crystal surfaces, so they thread through the body of the crystal and result in the extensive area of APBs, illustrated in a sketch shown in Figure 7a. The dislocation line direction is <110>, so the dislocations can exist along six <110> directions inside the single crystal. These extended defects have compressive, tensile and shear strains along <111>, <112>, <001>, and <110>. As a result, these strains would cause the Yb-tetrahedra to distort in different ways (Figure 7b). All four $Yb^{3+}$ move away from equilibrium positions in the tetrahedron. This deformed Yb-tetrahedron will make the delicate magnetic interactions in this geometry to deviate from the magnetic spin balance of the four $Yb^{3+}$ ions of a perfect lattice. The distorted Yb-tetrahedra from these extended defects, which are non-uniformly distributed inside the crystal, can potentially disturb the long-range $Yb^{3+}$ magnetic interactions, thus affect the low temperature physical properties of this compound, such as magnetic ground states. Since the EELS results did not find Ti valence change at the defects, there might not be an appreciable level of oxygen vacancies associated with these defects.

In addition to off-stoichiometric defects, which are realized by point defects or/and vacancies, the extended APBs and dissociated superdislocations can significantly affect the physical properties of the $Yb_2Ti_2O_7$ pyrochlore system. With the observation and characterization of these newly discovered defects, we provide a more comprehensive picture of the nature of existing defects in the FZ single crystals. Depending on the syntheses and growth conditions, even with stoichiometric materials, it's very likely that there are superdislocations present in some of the crystals. We postulate that these superdislocations and extended APB would render various magnetic properties of $Yb_2Ti_2O_7$ pyrochlore crystals because the dislocation density might be different in crystals with different growth conditions. They can cause variations in critical temperatures, and the presence or absence of static magnetic order at low temperatures. This seems to suggest that the broad phase transition reflected in specific heat could be explained by the non-homogeneous distribution of distorted Yb-tetrahedra in a three-dimensional arrangement. These line and planar defects would offer a natural explanation for a non-magnetic, disordered states in stoichiometric single crystals.

Although thermal annealing processes can potentially reduce dislocation densities, however, for the $Yb_2Ti_2O_7$ crystals, it might not be effective as it is difficult for the superdislocations to move due to their high energy and low mobility, where the extended complicated core structures would impede the dislocation motion. This assumption is supported by the fact that some crystals do not have any change of properties upon annealing[10]. Density functional theory calculations on these line defects are being carried out to understand the strain field effect on the four $Yb^{3+}$ spin interactions. Further studies are needed to explore these defects in depth both experimentally and theoretically to directly relate them to the various magnetic ground state of this material system. This work also suggests the importance of studying dislocation core structures in complex material systems.

## Methods

**Synthesis and crystal growth**. There are several types of crystals used in this study. A polycrystalline powder sample that has a white color was synthesized by mixing and grinding stoichiometric mixture of $Yb_2O_3$ (pre-dried at 980 °C) and $TiO_2$ together and calcining in air at 1400 °C for 40 h with an intermediate grinding. In this report, we looked at three single crystals grown by FZ and traveling solvent floating zone method (TSFZ) method. The colored crystals were grown at high temperatures (at the melting point and above) by the conventional floating zone (FZ) technique using two-mirror (Canon Machinery) and four-mirror (Crystal Systems Inc. FZ-T-4000-H-VII- VPO-PC) optical furnaces. The more yellowish single crystal was grown at a fast zoning rate of 25 mm/h under 5 bar oxygen pressure, while the light brownish color crystal was grown at a slow rate of 5 mm/h under 2 bar oxygen pressure. Regardless of the used furnace type, growth speed or gas pressure applied, this compound is not a fully congruently melting compound and both crystals are probably non-stoichiometric[9]. Colorless high quality stoichiometric single crystals were grown by the TSFZ method[9].

**X-ray diffraction and Magnetic measurement**. Powder X-ray diffraction data were collected on both powdered polycrystalline and FZ single-crystal samples at room temperature using Cu Kα1 radiation (λ=1.54051 Å). The data were analyzed using the Rietveld refinement method. Magnetic susceptibility was measured as a function of temperature using a SQUID magnetometer. For SQUID measurements, the sample was cooled in zero field to 2 K; then the susceptibility data was taken while warming the sample to 300 K in a 100 Oe magnetic field.

**TEM measurement** TEM samples were prepared by crushing samples in ethanol with a mortar and pestle. This ethanol suspension of crushed crystals was dropped with a pipette onto a 200 µm mesh TEM copper grid which is coated with carbon/formvar film from Ted Pella, Inc.. TEM sample prepared this way preserved the pristine and original quality of the crystals. TEM grid coated with carbon/formvar film instead of holey carbon film had to be used because sample charging occurred if the crystal pieces hanging over the hole. The charging on the sample surface caused broadening of the electron probe resulting in poor imaging resolution. STEM study was carried out on a probe-aberration-corrected, cold-field-emission JEM JEOL-ARM200cF at 200 kV using a JEOL HAADF-STEM detector. The STEM resolution is 0.078 nm, and energy resolution is 0.5 eV at full emission.

HAADF-STEM images were acquired using a condenser lens setting 7c, 30 µm condenser lens aperture, at a camera length of 8cm or 6 cm corresponding to an inner collection angle of 68 mrad and 90 mrad respectively, and an image scan speed of 32 $\mu$s/pixel. The beam convergent angle was 21 mrad. Core-loss EELS spectra from perfect area and defects were collected in STEM mode on a Gatan GIF Quantam963 while monitoring the exact location from live STEM image. The core-loss spectra were collected together with zero loss peak using DualEELS at 1.5 cm camera length with a 2.5 mm slit and an energy dispersion of 0.1 eV. Each core-loss spectrum was acquired by 2s and the final spectrum shown was the sum of about 10 individual spectra which had background subtracted and plural scattering removed.

Theoretical calculations have been carried out using xHREM$^{TM}$ STEM simulation software package. An artificial unit cell contains one Yb and one Ti with 3.645 Å apart along the electron beam direction was constructed so that the atom distance is the same as in the $Yb_2Ti_2O_7$ structure. The calculation conditions used are: image collection angle from 59 -159 mrad, and Debye-Waller factors for Yb and Ti are 0.175 and 0.313 Å$^2$ respectively. The probe was focused on the sample surface.

## Acknowledgements

This work was performed at the National High Magnetic Field Laboratory of Florida State University, which is supported by National Science Foundation Cooperative Agreement No. DMR-1157490, DMR-1644779, and the State of Florida. H.D.Z. thanks for the support from NSF-DMR with grant number NSF-DMR-1350002. Q.H. thanks for the support from the Go students program of ORNL. The Institute for Quantum Matter is supported by the Department of Energy (DOE), Office of Basic Energy Sciences, Division of Materials Sciences and Engineering, under award DE-SC0019331.

## Author contributions

Y.X. supervised and designed the experiments. Y.X. collected the TEM data together with Z.S.. H.Q., H.D.Z and S.M.K. grew the crystals. H.Q. and H.D.Z did X-ray diffraction experiments and magnetic property measurements on powder and FZ crystal samples prepared at the University of Tennessee. Y.X. performed TEM data analysis and wrote the paper. Z.S. carried out EELS data processing and theoretical calculations. All authors reviewed and commented on the paper.

## Additional Information

Supplementary information accompanies this paper.

Competing Interests statement: The authors declare no competing financial AND non-financial interests.

**Figure legends**

Figure 1 (a) HAADF-STEM image of one crystal piece from a TSFZ single crystal sample viewed along [110]. (b) Normalized atomic column cross-section intensity map of (a) using Absolute Integrator software by Lewys Jones.[23] (c) Bar graphs of measured Yb column intensity and Ti column intensity from the boxed atomic row in (a). The Ti column intensity is varied about 5% for the perfect region. (d) HAADF-STEM image of one crystal piece from a stoicheometric polycrystalline sample viewed along [110]. (e) Normalized atomic column cross-section intensity map of (d). (f) Bar graphs of measured Yb column intensity and Ti column intensity from the boxed atomic row in (d).

Figure 2 (a) HAADF-STEM image of one crystal piece from a colored FZ single-crystal sample viewed along [110]. (b) Normalized atomic column cross-section intensity map of (a). (c)-(f) Bar graphs show the measured intensity of Ti columns, and Yb columns from the atomic rows indicated by the red boxes in (a). The Ti columns with substitution Yb atoms are indicated by arrows.

Figure 3    A superdislocation with Burgers vector ¼ [1$\bar{1}$2]. (a) The dissociated superdislocation with two partial dislocations and an APB with three Burgers circuits. The projected Burgers vectors are shown by the red arrows.  (b) Inverse FFT image using $\mathbf{g} = \pm(2\bar{2}2)$; inset: FFT diffraction with $\mathbf{g}$ circled. (c)  Inverse FFT image using $\mathbf{g} = \pm (2\overline{2}\overline{2})$; inset: FFT diffraction with $\mathbf{g}$ circled. (d) An enlarged HAADF-STEM image of a 1/8[1$\bar{1}$2] partial dislocation. (e) Inverse FFT image using $\mathbf{g} = \pm(2\bar{2}2)$. The (1$\bar{1}$1) atomic plane has a kink/step at the core due to a screw component. (f)-(h) $\varepsilon_{xx}$ along [1$\bar{1}$0], $\varepsilon_{yy}$ along [00$\bar{1}$], and  $\varepsilon_{xy}$ strain maps of (a); the location of the two extra atomic planes of the dislocation cores indicated by lines.

Figure 4    Atomic structures of APBs and its strain map: (a) HAADF-STEM image of a perfect lattice and intensity line profiles of Ti-Yb/Ti and Yb/Ti atomic rows from the red boxes. (b) A defect area and intensity line profiles showing little change of atomic column intensity. (c) HAADF-STEM image of the same type of APB in an area. The APB is on {111} with seven atomic planes wide. (d) Phase image of (c), calculated using diffraction $\mathbf{g} = (1\bar{1}1)$, clearly

showing phase reverse across the defect. Strain field by GPA analysis shows (e) $\epsilon_{xx}$ [1$\bar{1}\bar{2}$]; (f) $\epsilon_{yy}$ along [1$\bar{1}$1] and (g) shear strain $\epsilon_{xy}$.

Figure 5    A partial dislocation from a dissociated superdislocation with a projected Burgers vector ¼ <110>. (a) HAADF-STEM image of one partial dislocation bounding an APB. Yellow frame is the Burgers circuit. The projected Burgers vector is shown by the red arrow. (b) and (c) Inverse FFT images using **g** = ±(2$\bar{2}$2) and ±($\bar{2}$22) show two extra planes at the partial dislocation core. (d) FFT filtered HAADF-STEM image of the core structure of the partial dislocation. (e) Phase map superimposed onto (a). The phase changed from –π to π across the (001) APB. (f)-(h) Strain map of (a): (f) $\epsilon_{xx}$ along [1$\bar{1}$1]; (g) $\epsilon_{yy}$ along [1$\bar{1}\bar{2}$]; (h) shear strain $\epsilon_{xy}$

Figure 6    (a) Ti $L_{2,3}$ core-loss EELS spectra from perfect region and from APB. They are normalized to the first peak of the doublet of $L_3$. The second peak of the $L_2$ doublet is slightly higher for APB. (b) O-K edge core-loss EELS spectra on and off APB.

Figure 7    (a) Schematic diagram to illustrate the distribution of superdislocations inside a single crystal. The six <110> direction is indicated by arrows. (b) $Yb^{3+}$ ions displacement away from equilibrium position in a Yb-tetrahedron due to strain indicated by the arrows.

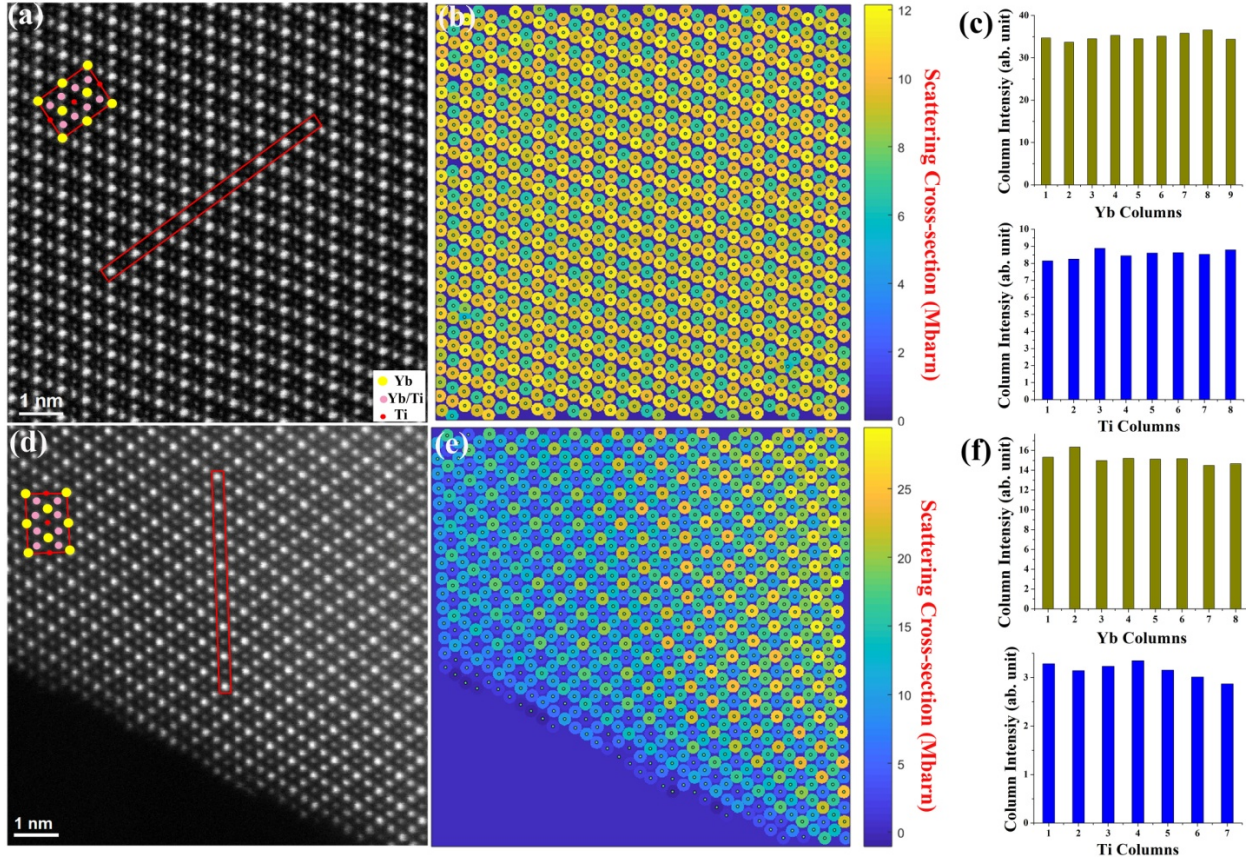

**Figure 1**

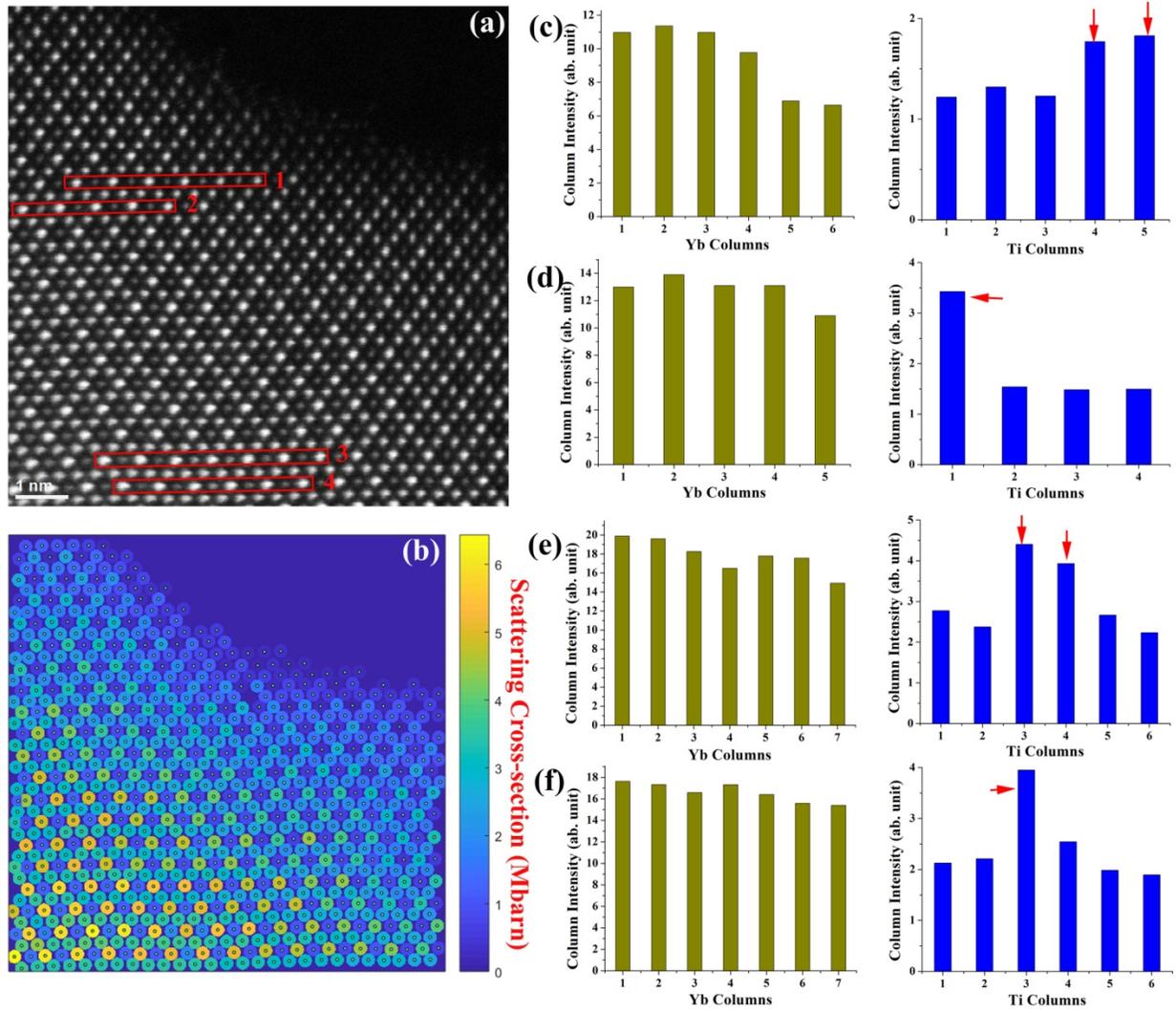

Figure 2

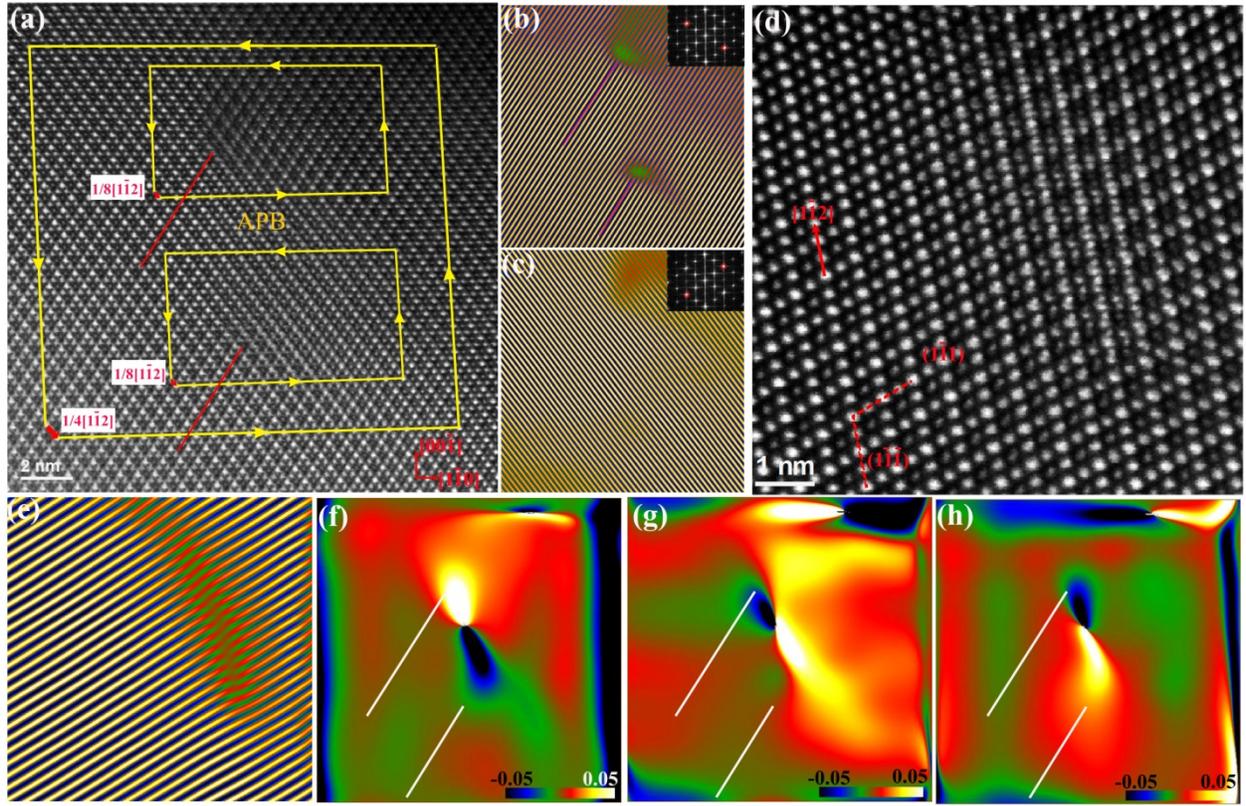

**Figure 3**

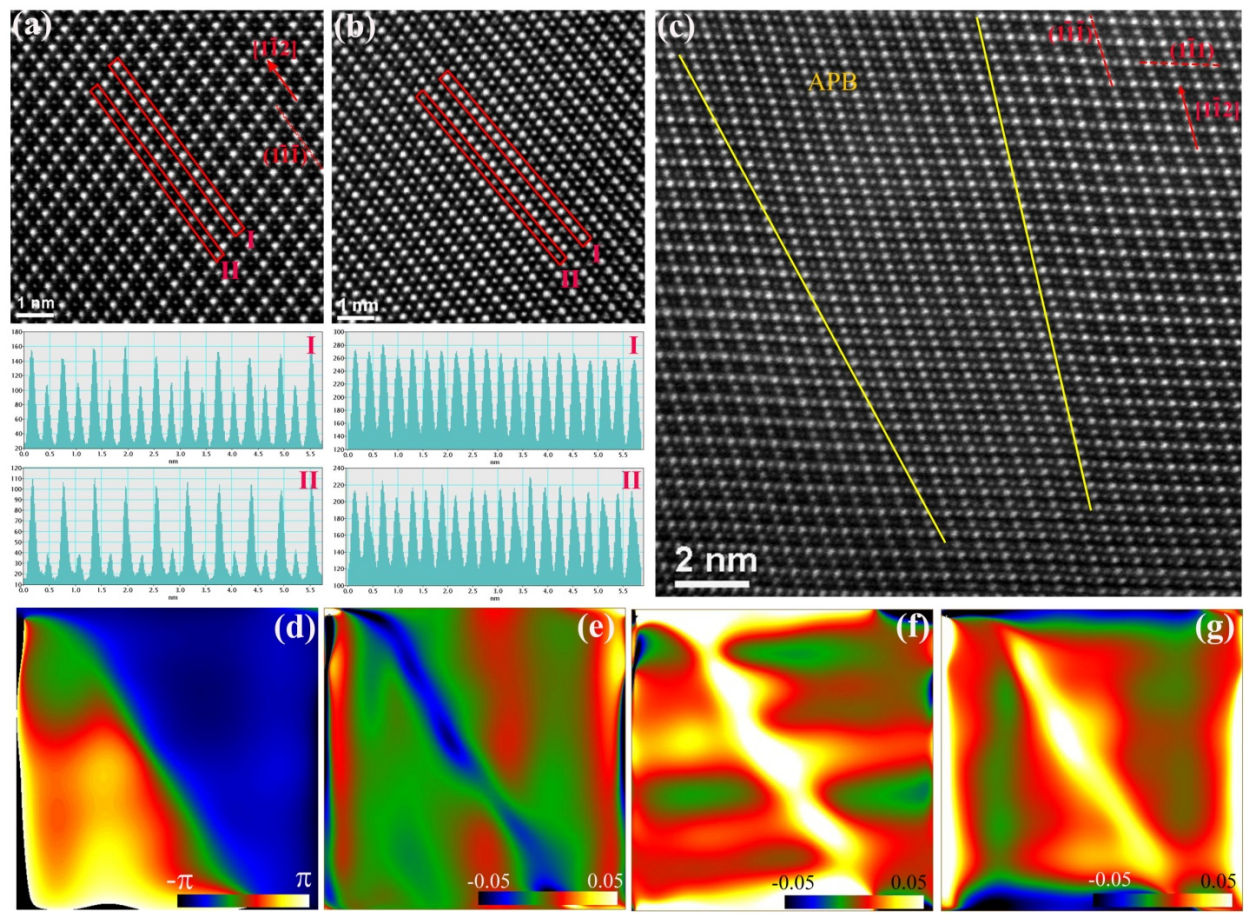

**Figure 4**

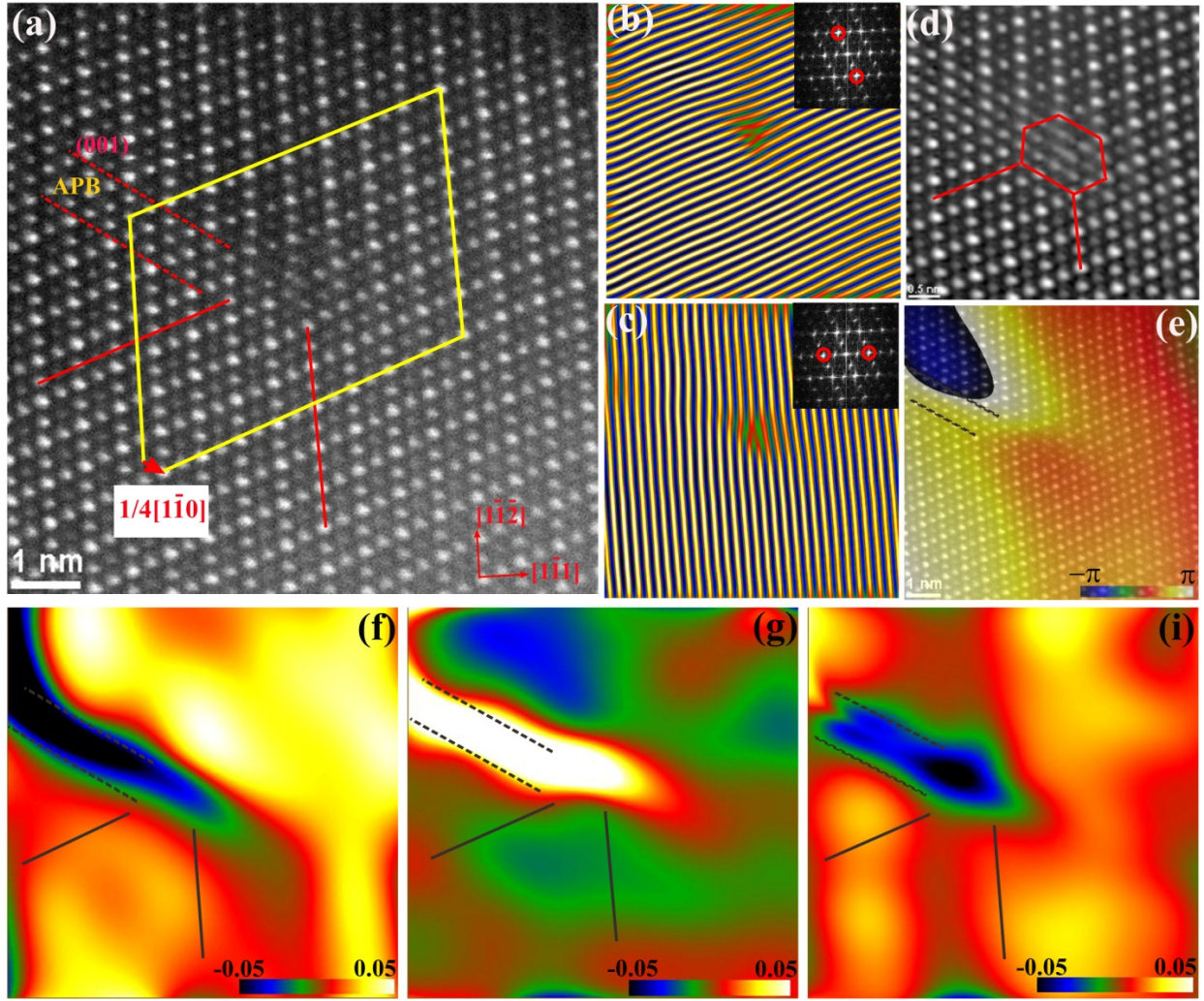

**Figure 5**

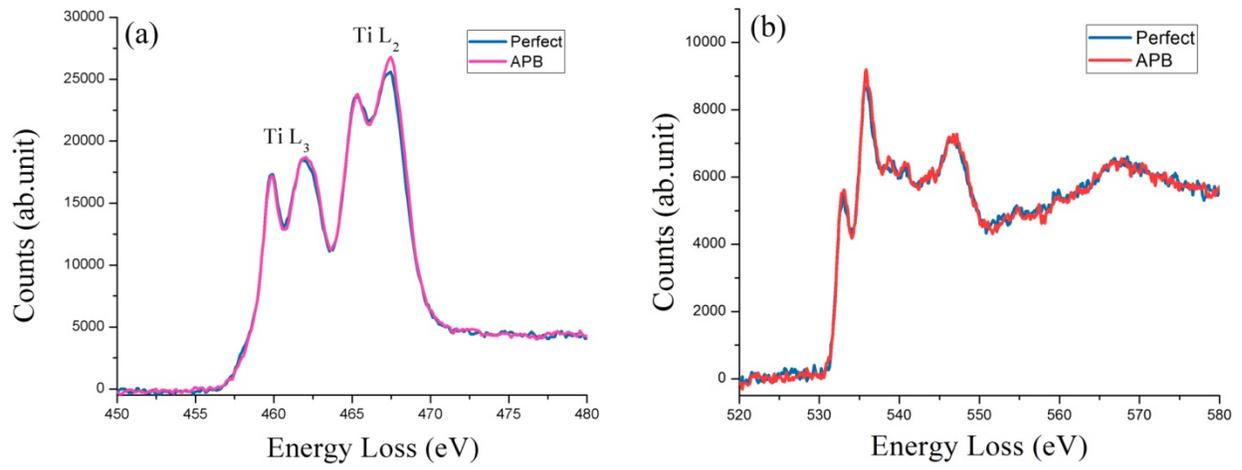

**Figure 6**

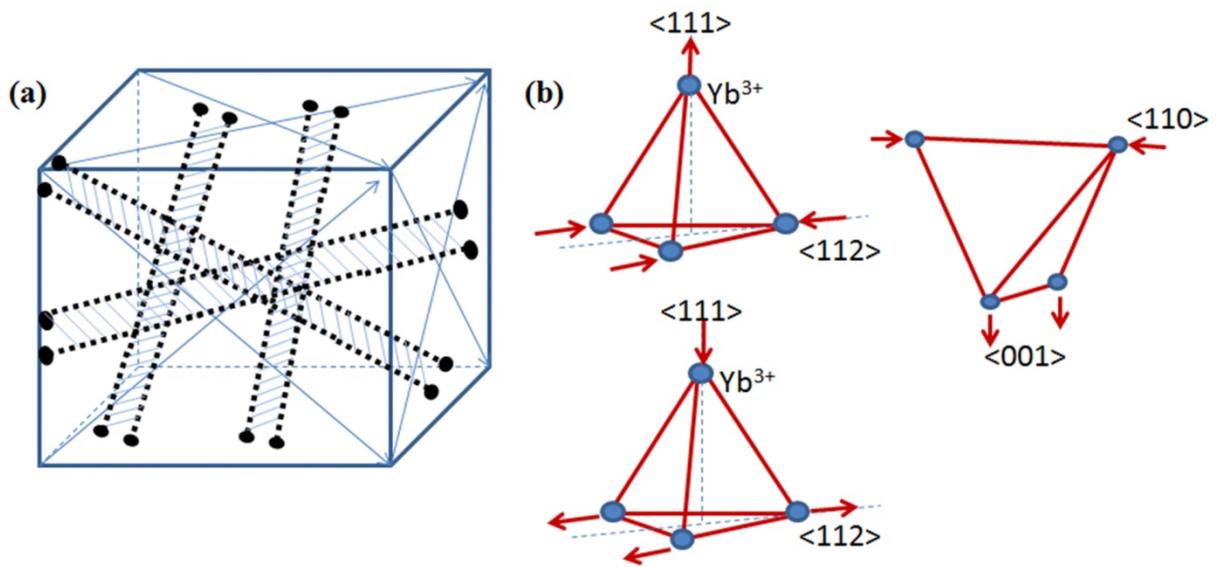

**Figure 7**

**Supplementary materials**

**Superdislocations and point defects in pyrochlore Yb$_2$Ti$_2$O$_7$ single crystals and implication on magnetic ground states**

Zahra Shafieizadeh[1,+], Yan Xin[2,+,*], Seyed M. Koohpayeh[3], Qing Huang[4], & Haidong Zhou[4]

[1] Department of Physics, Florida State University, Tallahassee, FL 32311, USA

[2] National High Magnetic Field Laboratory, Florida State University, Tallahassee, FL 32310, USA

[3] Institute for Quantum Matter, Department of Physics and Astronomy, Johns Hopkins University, Baltimore, MD 21218, USA

[4] Department of Physics and Astronomy, University of Tennessee, Knoxville, TN 37996, USA

[+] These two authors contributed equally to this work.

[*]Correspondence and requests for materials should be addressed to Y.X. (email: xin@magnet.fsu.edu)

Contact information:

Dr. Yan Xin

National High Magnetic Field Laboratory

Florida State University

1800 E. Paul Dirac Drive, Tallahassee, FL 32310

USA

Phone: (850) 644-1529

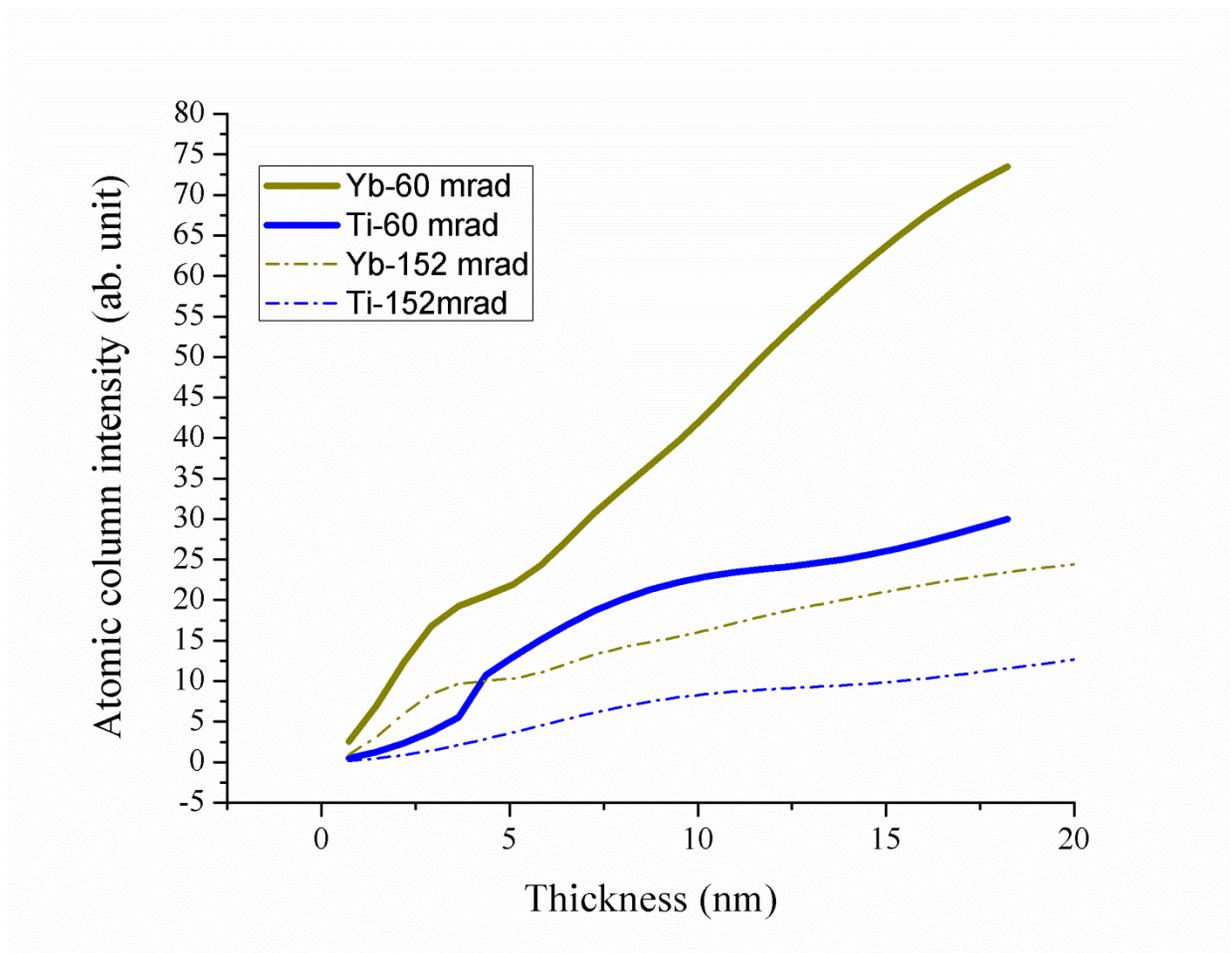

Figure S1 Yb and Ti atomic column intensity vs sample thickness with different inner collection-angle. The solid curve corresponds to the experimental condition used. This calculation confirms that the intensity difference for Yb and Ti columns is mainly due to the atomic number effect rather than diffraction contrast effect.

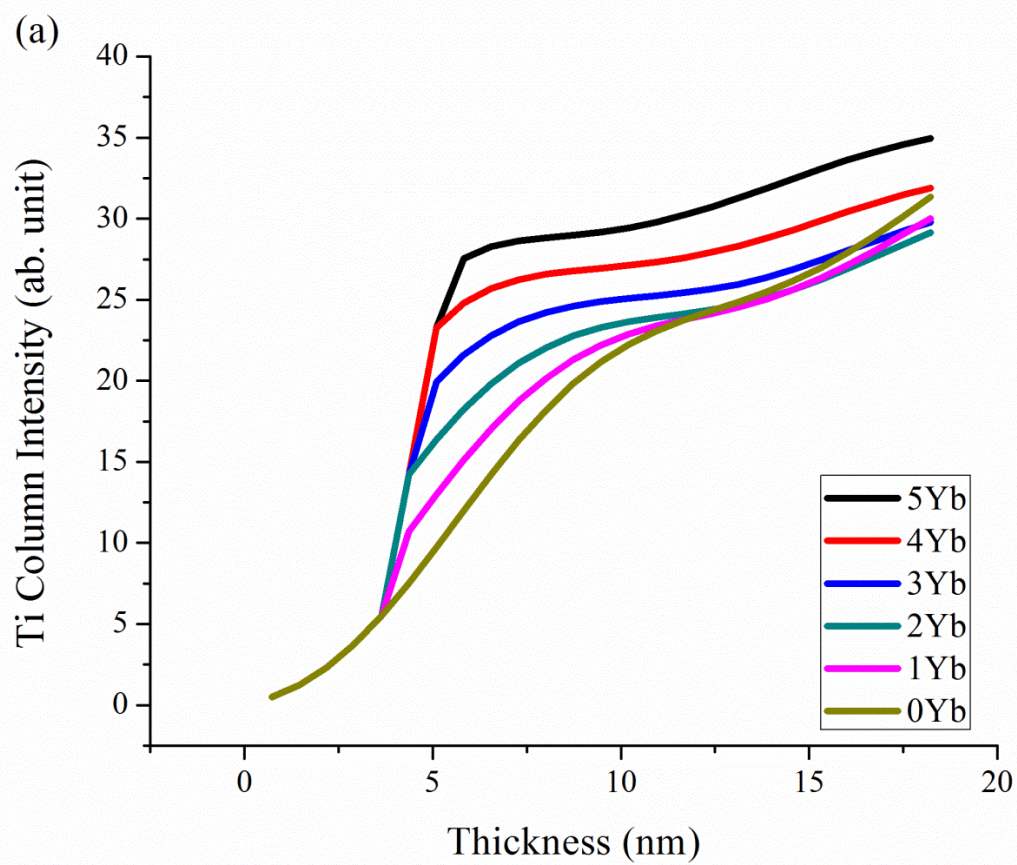

Figure S2 (a) Ti column intensity vs thickness curve for one to five substitution Yb atoms at 3.65 nm below surface;

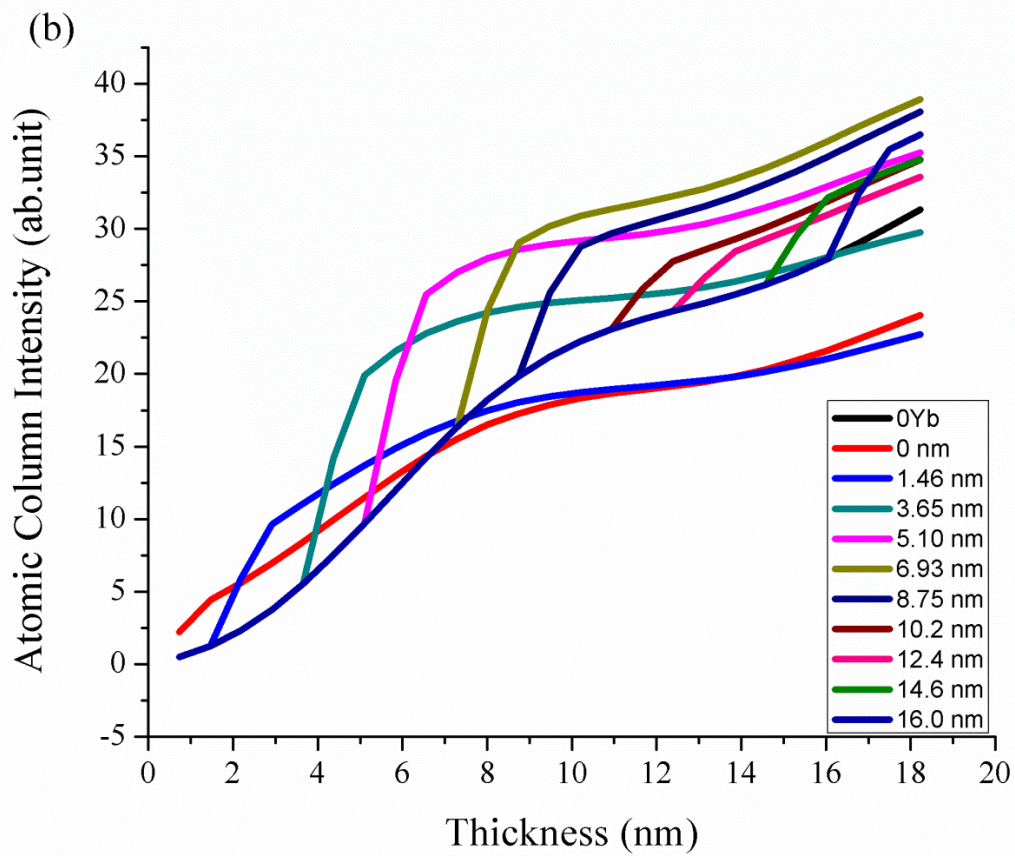

Figure S2 (b) Ti atomic column intensity vs thickness curve for 3 substitution Yb atoms inside Ti column at different depths below sample surface.

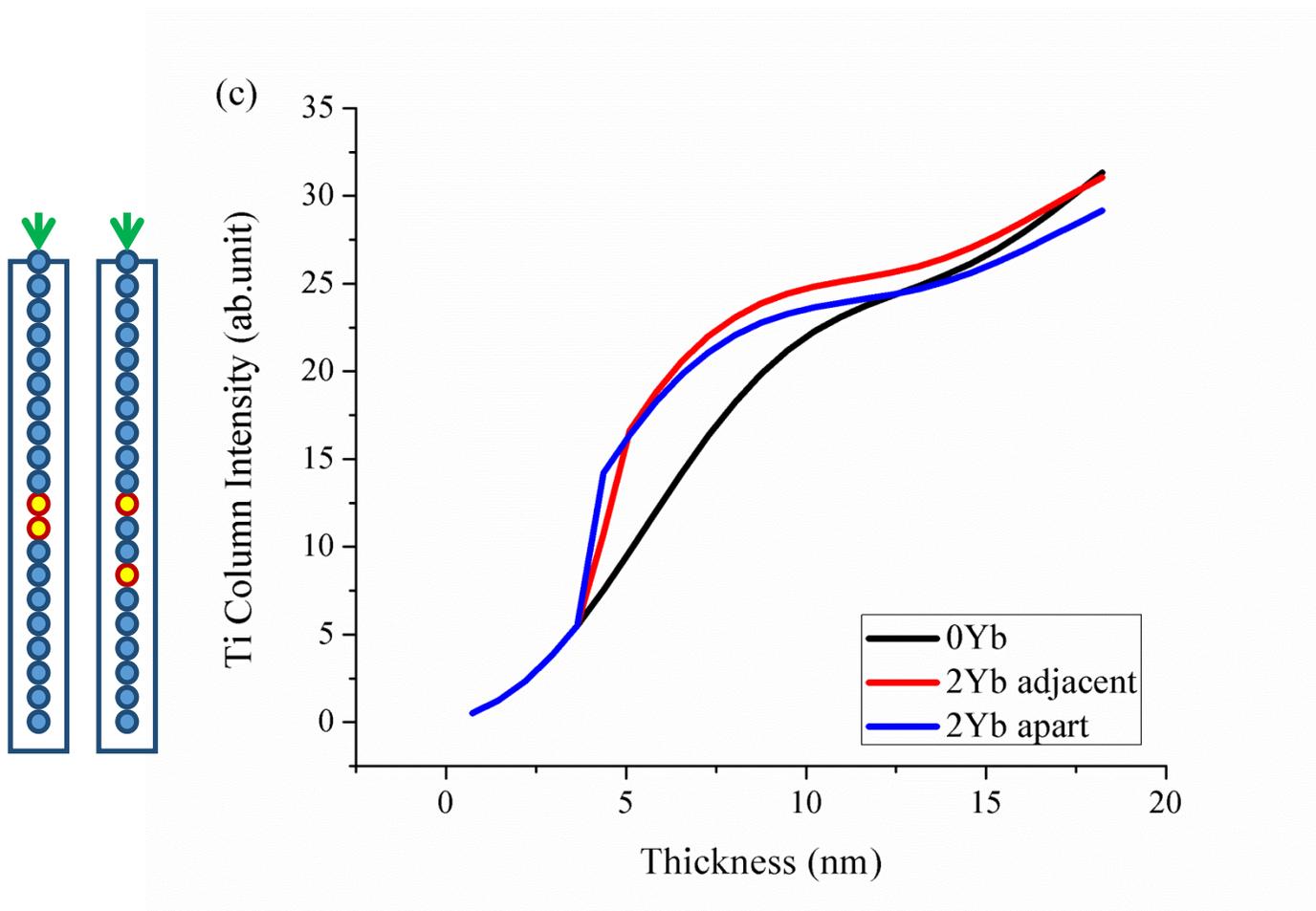

Figure S2 (c) Ti atomic column intensity vs thickness curve for 2 substitution Yb atoms inside Ti column adjacent to each other and apart as illustrated in sketch on the left;

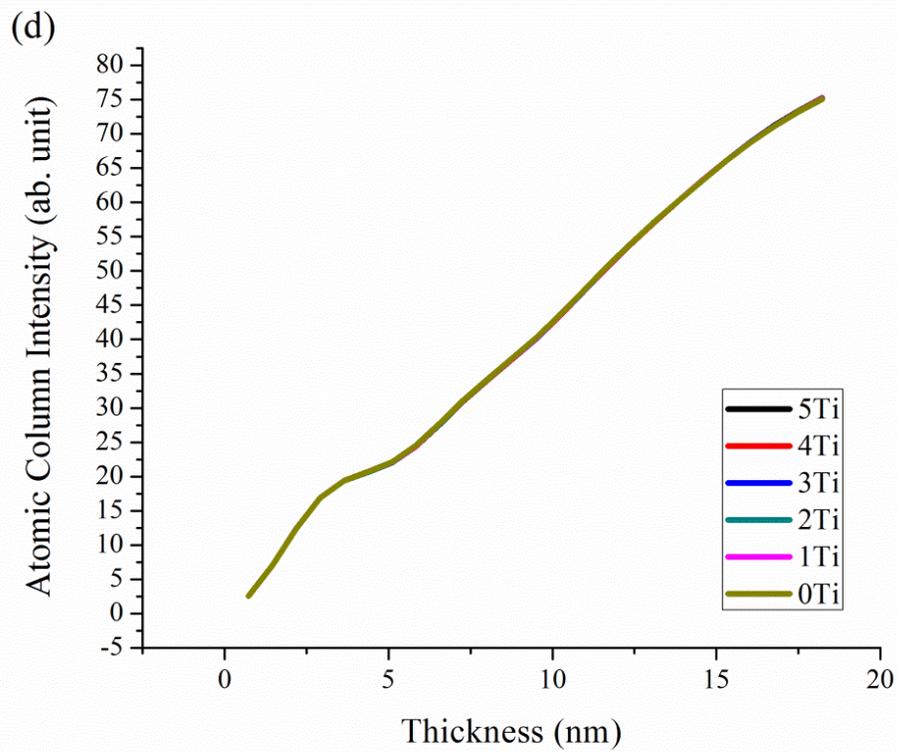

Figure S2 (d) Yb atomic column intensity vs thickness curve of one to five Ti substitution into Yb columns. All five curves overlapping on top of each other.